\documentclass[12pt]{article}

\ifx\pdfoutput\undefined
\usepackage[dvips,bookmarks]{hyperref}
\else
\usepackage{hyperref}
\fi
\hypersetup{colorlinks=false,bookmarksopen,bookmarksnumbered,citecolor=blue,
   pdfstartview=FitH}

\usepackage{latexsym}
\usepackage{amssymb,amsfonts,amsmath}
\usepackage{graphicx} 
\usepackage{indentfirst}
\usepackage{bbm}
\usepackage{amssymb}
\usepackage{verbatim}
\usepackage{amsmath, amsthm,amssymb}
\usepackage{mathrsfs}
\usepackage{hyperref}
\usepackage{amsfonts}
\usepackage{dsfont}
\usepackage{slashed, tensor}
\usepackage{booktabs}
\usepackage{graphicx}
\usepackage{mathrsfs}
\parskip=\medskipamount

\newcommand {\cB}{{\cal B}}

\newcommand {\cD}{{\cal D}}

\newcommand {\cF}{{\cal F}}

\newcommand {\cH}{{\cal H}}

\newcommand {\cL}{{\cal L}}

\newcommand {\cN}{{\cal N}}

\newcommand {\cV}{{\cal V}}
\newcommand {\cW}{{\cal W}}

\def\a{\alpha}
\def\b{\beta}

\def\d{\delta}

\def\g{\gamma}
\def\G{\Gamma}

\def\k{\kappa}
\def\l{\lambda}

\def\o{\omega}
\def\p{\pi}
\def\q{\theta}
\def\r{\rho}

\def\t{\tau}

\def\x{\xi}
\def\z{\zeta}

\def\L{\Lambda}

\def\tr{{\rm tr}}

\def\ri{{\rm i}}
\def\re{{\rm e}}

\newcommand{\ad}{{\dot{\alpha}}}                           %new
\newcommand{\ve}{\varepsilon}                            %new
\newcommand{\pa}{\partial}                           %new
\newcommand{\hf}{\frac12}
\newcommand{\be}{\begin{equation}}
\newcommand{\ee}{\end{equation}}
\newcommand{\bea}{\begin{eqnarray}}
\newcommand{\eea}{\end{eqnarray}}
\newcommand{\non}{\nonumber}
\newcommand{\ba}{\begin{array}}
\newcommand{\ea}{\end{array}}

\newcommand{\1}{{\underline{1}}}
\newcommand{\2}{{\underline{2}}}

\def\double #1{#1{\hbox{\kern-2pt $#1$}}}

\newcommand{\bsubeq}{\begin{subequations}}
\newcommand{\esubeq}{\end{subequations}}

\newcommand{\eps}{{\ve}}

\newcommand{\rd}{\mathrm d}
\numberwithin{equation}{section}
\begin{document}
%%%%%%%%%%%%%%%%
%%%%%%%%%%%%%%%%

\begin{titlepage}
\begin{flushright}
November, 2015 \\
\end{flushright}
\vspace{5mm}

\begin{center}
{\Large \bf 
The anomalous current multiplet in 6D minimal supersymmetry
}
\\ 
\end{center}

\begin{center}

{\bf
Sergei M. Kuzenko${}^{a}$, Joseph Novak${}^{b}$ 
and Igor B. Samsonov${}^{a,}$\footnote{On leave from Tomsk Polytechnic University, 634050
Tomsk, Russia.}
} \\
\vspace{5mm}

\footnotesize{
${}^{a}${\it School of Physics M013, The University of Western Australia\\
35 Stirling Highway, Crawley, W.A. 6009, Australia}}  
~\\
\vspace{2mm}
\footnotesize{
${}^{b}${\it Max-Planck-Institut f\"ur Gravitationsphysik, Albert-Einstein-Institut,\\
Am M\"uhlenberg 1, D-14476 Golm, Germany.}
}
\vspace{2mm}
~\\
\texttt{sergei.kuzenko,igor.samsonov@uwa.edu.au, joseph.novak@aei.mpg.de}\\
\vspace{2mm}

\end{center}

\begin{abstract}
\baselineskip=14pt

For supersymmetric gauge theories with eight supercharges in four, five and six 
dimensions, a conserved current belongs to the linear multiplet. 
In the case of six-dimensional $\cN=(1,0) $ Poincar\'e supersymmetry, 
we present a consistent deformation of the linear multiplet which describes 
chiral anomalies.  This is achieved by developing a superform formulation 
for the deformed linear multiplet. In the abelian case, we compute 
a nonlocal effective action generating the gauge anomaly.
\end{abstract}

\vfill

\vfill
\end{titlepage}

\newpage
\renewcommand{\thefootnote}{\arabic{footnote}}
\setcounter{footnote}{0}

%%%%%%%%%%%%%%%%%%%%%%%%%%%%%%%%%%%%%%%%%%%%%%%%%%%%%%

\tableofcontents
\vspace{1cm}
\bigskip\hrule

%%%%%%%%%%%%%%%%%%%%%%%%%%%%%%%%%%%%%%%%%%%%%%%%%%%%%%

\allowdisplaybreaks

\section{Introduction}

For supersymmetric gauge theories with eight supercharges in four, five and six 
dimensions, a conserved current belongs to the so-called linear multiplet \cite{BS}.
This  multiplet is described by a real SU(2) triplet superfield, $L^{ij}=L^{(ij)}$
and  $\overline{L^{ij}}=L_{ij} := \ve_{ik} \ve_{jl} L^{kl}$, 
 subject to the constraint
\be 
D_\a^{(i} L^{jk)} =0~,
\label{1.1}
\ee
with $\a$ the four-component spinor index.
Conserved current multiplets in $\cN=3$ and $\cN=4$ supersymmetric field theories
in three dimensions have similar structure \cite{BKS}. 

It is well known that the 
four-dimensional $\cN=2$ supersymmetric Yang-Mills theories 
are non-chiral. There are no chiral fermions in five dimensions. 
Thus all supersymmetric gauge theories with eight supercharges 
in four and five dimensions are  anomaly-free.
However, in six dimensions all irreducible matter representations of $\cN=(1,0)$
supersymmetry (the hypermultiplet, the vector multiplet and the tensor multiplet) 
as well as the supergravity multiplet contain chiral fermions.
Moreover, the tensor and supergravity multiplets also contain 
chiral bosonic fields, which are gauge two-forms with (anti) self-dual field strengths.
These features imply the existence of numerous chiral $\cN=(1,0)$ supersymmetric 
gauge theories. That is why the classification and structure of anomaly-free 
6D supersymmetric theories were  thoroughly studied  in the 1980s
at the component level,  
see \cite{SalamSezgin} for a review. 

In the presence of anomalies, 
the 6D conservation equation \eqref{1.1} turns into 
\begin{subequations}
\bea 
D_\a^{(i} L^{jk)} = A_\a^{ijk}~,
\label{1.2a}
\eea
for some superfield $A_\a^{ijk}=A_\a^{(ijk)}$ constrained by\footnote{The superfield $A_\a^{ijk}$ 
subject to the constraint (1.2b) corresponds to a closed six-form, see \cite{ALR}.}
\bea
D_{(\a}^{(i} A_{\b)}^{jkl)} = 0 ~.
\label{consistency}
\eea
\end{subequations}
It was conjectured in  \cite{HU} that 
 $A_\a^{ijk}$ has the following structure:
 \bea
 A_\a^{ijk} = D_{\a l} A^{ijkl}~, \qquad D_{\a }^{(i} A^{jklm)} =0~.
 \label{1.3}
\eea
This led the authors of \cite{HU} to conclude that the anomalous 
current multiplet is a 6D relaxed hypermultiplet \cite{HST}.
A decade later, this conclusion was re-considered by Howe and Sezgin \cite{HS}
who studied the $\cN=(1,0)$ Yang-Mills multiplet coupled to a tensor multiplet. 
By analysing the one-loop corrected equations of motion, 
they found an expression for $A_\a^{ijk}$
that was incompatible with \eqref{1.3}.

In this paper we will argue that the anomalous 6D $\cN=(1,0)$ 
current multiplet $L^{ij}$ obeys the equation
\bea 
D_\a^{(i} L^{jk)} = \k \,\ri \,\eps_{\a\b\g\d}  \cW^{i \b } \cW^{j \g } \cW^{k \d } ~,
\label{1.4}
\eea
where $\cW^{i \a}$ is the field strength of an abelian vector multiplet 
\cite{HSierraT,Koller}, 
and $\k$ a real parameter. 
As we will show, it follows from this equation that there exists a current $j^a$ at the component level satisfying $\partial_a j^a \propto \eps^{abcdef} f_{ab} f_{cd} f_{ef}$, with $f_{ab}$ the component gauge invariant field strength.
Equation \eqref{1.4} can be shown to be superconformal
assuming $L^{ij} $ is a primary operator of dimension 4. 
On the contrary,  it may be shown that the ``relaxed hypermultiplet'' \eqref{1.3} 
does not describe a representation of the superconformal group.

Equation \eqref{1.4} appeared in \cite{HS} in the context of the model 
describing the 6D $\cN=(1,0)$ Yang-Mills multiplet coupled 
to the tensor multiplet.\footnote{The analysis in \cite{HS} is rather inconclusive, 
as may be deduced from the final comments given in that paper. The authors state that
 ``in order to capture the full supermultiplet structure
 of the equations of motion plus anomalies, one should relax 
the ``relaxed hypermultiplet'' even further.''}
Here we will argue that the constraint has a universal nature.\footnote{In \cite{HS} Howe and Sezgin 
considered the non-abelian vector multiplet. Eq. \eqref{1.4} 
admits  a straightforward generalisation and we leave discussion of it to 
section \ref{section2}.} 
The simplest anomalous $\cN=(1,0)$ supersymmetric theory is a hypermultiplet
coupled to a U(1) vector multiplet. The hypermultiplet contains a single chiral fermion
interacting with the gauge field. The corresponding gauge anomaly is well known
\cite{Frampton,A-GW,ZuminoZee,Leutwyler}
\bea
\partial^a j_a = 
- \frac1{384\pi^3}\varepsilon^{abcdef}f_{ab}f_{cd}f_{ef}~.
\label{1.5}
\eea
It corresponds to eq. \eqref{1.4} with $\k=\frac1{96\pi^3}$ and $j_a$ a component field of 
$L^{ij}$ (defined by eq. \eqref{2.399}).

This paper is organised as follows. Section 2 gives a superform formulation for the anomalous current multiplet. In particular, 
the consistency of the  constraint (1.4) is demonstrated. In section 3 we 
compute the nonlocal effective action generating the chiral anomaly. 
We also discuss how the problem of chiral anomalies should be addressed
in the frameworks of the harmonic and the projective superspace approaches. 
Section 4 is devoted  to an alternative description of the anomalous current multiplet.
Concluding comments are given in section 5. The paper also includes two technical appendices. Appendix A is devoted to a brief review of the three prepotential formulations for the 6D $\cN=(1,0) $ vector multiplet. 
Appendix B contains the technical details concerning the 
derivation of the anomalous effective action.

%%%%%%%%%%%%%%%%%%%%%%%%%%%%%%%%%%%%%%%%%%%%%%%%%%%%%%
%%%%%%%%%%%%%%%%%%%%%%%%%%%%%%%%%%%%%%%%%%%%%%%%%%%%%%

\section{Superforms and the anomalous current multiplet}\label{section2}

In this section we present a superform construction for the anomalous current. 
We work in standard 6D $\cN=(1,0) $ Minkowski superspace 
parametrised by coordinates $z^A = (x^a, \q^\a_i)$. 
Our 6D notation and conventions correspond to \cite{LT-M}.
In particular, the flat-superspace covariant derivatives are 
denoted by $D_A = (\pa_a, D_\a^i)$, and the dual basis of one-forms 
is denoted by $E^A$, such that $\rd = \rd z^M \pa_M = E^A D_A$. 

%%%%%%%%%%%%%%%%%%%%%%%%%%%%%%%%%%%%%%%%%%%%%%%%%%%%%%
%%%%%%%%%%%%%%%%%%%%%%%%%%%%%%%%%%%%%%%%%%%%%%%%%%%%%%

\subsection{The vector multiplet}

In this subsection we review the superspace formulation for the 6D
$\cN=(1,0)$ Yang-Mills supermultiplet following \cite{HST}.
To describe a non-abelian vector multiplet, the covariant derivative $D_A$  
has to be replaced with a gauge covariant one, 
\bea
\cD_A := D_A 
+ \ri \cV_A
~.
\label{SYM-derivatives}
\eea
Here the  gauge connection one-form $\cV = E^A \cV_A$ 
takes its values in the Lie algebra 
of the Yang-Mills gauge group $G_{\rm YM}$. 
The covariant derivative algebra is
\bea
[{\cD}_A, {\cD}_B \} 
&=&
 T_{AB}{}^C{\cD}_C
	+ \ri \cF_{AB} \ ,
\label{gauged-algebra}
\eea
where the only non-vanishing torsion is
\be T_\a^i{}_\b^j{}^c = - 2 \ri \eps^{ij} (\g^c)_{\a\b}
\ee
and $\cF_{AB}$ corresponds 
to the gauge covariant field strength two-form. The covariant derivatives and field strength may be written in a 
coordinate-free way as follows
\be \cD = \rd + \ri \cV \ , \quad \cF = \rd \cV - \ri \cV \wedge \cV \ , \label{CovDer}
\ee
where
\be \cD := \rd z^{A} \cD_{A} \ , \quad \cV := \rd z^{A} \cV_{A} \ , \quad 
\cF := \hf \rd z^{B} \wedge \rd z^{A} \cF_{AB} \ .
\ee
The field strength $\cF_{AB}$ 
satisfies the Bianchi identity
\be \cD \cF = \rd \cF + \ri \cV \wedge \cF - \ri \cF \wedge \cV = 0  
\quad \Longleftrightarrow \quad
\cD_{[A} \cF_{B C \} } - T_{[A B}{}^{D} \cF_{|D|C \} } = 0 \ .
\ee
The Yang-Mills gauge transformation acts on the gauge covariant 
derivatives $\cD_A$ and a matter  superfield $U$ (transforming 
in some representation of the gauge group) 
as
\be 
\cD_A ~\rightarrow~ \re^{\ri  \t} \cD_A \re^{- \ri \t } , 
\qquad  U~\rightarrow~ U' = \re^{\ri  \t} U~, 
\qquad \t^\dag = \t \ ,
\label{2.2}
\ee
where the Hermitian gauge parameter ${\t} (z)$ takes its values in the Lie algebra 
of $G_{\rm YM}$. This implies that the gauge connection and field strength transform as follows
\be
\cV \rightarrow  \re^{\ri \t } \,\cV \,\re^{- \ri \t} - \ri \re^{\ri \t } \,\rd \re^{- \ri \t } \ , 
\quad \cF \rightarrow \re^{\ri \t } \,\cF \,\re^{- \ri \t} \ .
\ee

Some components of the field strength have to be constrained 
in order to describe an irreducible multiplet. 
Upon constraining the lowest mass dimension component of the field strength tensor as 
\bsubeq \label{YMsuperformF}
\bea 
\cF_\a^i{}_\b^j = 0 \ , \quad \cF_a{}_\b^j = (\g_a)_{\b \g} \cW^{j\g} \ ,
\eea
the remaining component 
is completely determined to be
\bea
\cF_{ab} = - \frac{\ri}{8} (\g_{ab})_\b{}^\a \cD_\a^k \cW^\b_k \ ,
\eea
\esubeq
where the superfield $\cW^{i\a }$ 
obeys the Bianchi identities
\bea 
\cD_\g^k \cW^\g_k = 0 \ , \quad \cD_\a^{(i} \cW^{j)\b }
= \frac{1}{4} \d_\a^\b \cD_\g^{(i} \cW^{j)\g }
\label{vector-Bianchies} \ .
\eea
The vector indices of  $\cF_{ab}$ can be converted into spinor ones as follows:
\be 
\cF_\a{}^\b := -\frac{1}{4} (\g^{ab})_\a{}^\b \cF_{ab} 
= - \frac{\ri}{4} \Big( \cD_\a^k \cW^\b_k - \frac{1}{4} \d_\a^\b \cD_\g^k \cW^\g_k \Big)
= - \frac{\ri}{4} \cD_\a^k \cW^\b_k ~.
\ee

It is convenient to introduce the following superfield:
\begin{align}
X^{ij} := \frac{\ri}{4} \cD^{(i}_\g \cW^{j)\g } \ .
\end{align}
The superfields $\cW^{i\a}$,  $X^{ij}$
and $\cF_\a{}^\b $ 
satisfy the  useful identities:
\bsubeq \label{VMIdentities}
\bea
\cD_\a^i \cW^{j\b }
&=& 
- \ri \d_\a^\b X^{ij} - 2 \ri \eps^{ij} \cF_\a{}^\b \ , \\
\cD_\a^i  \cF_\b{}^\g
&=&
- \cD_{\a\b} \cW^{i\g}
- \d^\g_\a  \cD_{\b\d}  \cW^{i\d }
+ \frac{1}{2} \d_\b^\g  \cD_{\a \d} \cW^{i\d }
\ , \\
\cD_\a^i  X^{jk}
&=&
2 \eps^{i(j} \cD_{\a\b}  \cW^{k)\b }
\ .
\eea
\esubeq

The above identities indicate how to define the independent component fields 
contained in $\cW^{i \a}$. They may be
defined as follows:
\begin{align} \label{VMcompDef}
\l^{i\a } := \cW^{i\a } | \ , \quad f_\a{}^\b := \cF_\a{}^\b | \ , 
\quad y^{ij} := X^{ij} | \ ,
\end{align}
where the bar projection of a superfield $U(z) =U(x,\q)$ is defined by 
the standard rule $U| := U(x,\q)|_{\q = 0}$. The component gauge field is defined by 
$v_a := \cV_{a}|$ and is related to the component field strength $f_{ab}$ as follows
\be
f_{ab} = 2   \partial_{[a} v_{b]} + \ri \big[v_{a} ,v_{b } \big] \ .
\ee
It is seen that the vector multiplet consists of the following component fields: 
$\l^{i\a }$, $v_a$ and $y^{ij}$. 

The supersymmetry transformations of the fields 
$\l^{i\a }$, $v_a$ and $y^{ij}$
may be obtained by evaluating 
the component projection of the identities \eqref{VMIdentities}. This gives
\bsubeq \label{VMSUSY}
\begin{align}
\d_\xi \l^{i\a } &= - \ri \xi^\a_j y^{ij} + 2 \ri \xi^{\b i} f_\b{}^\a \ , \\
\d_{\xi} y^{ij} &= - 2 \xi^{\a (i} \cD_{\a\b} \l^{j)\b }\ , \\
\d_{\xi} v_a &= \xi^\b_j \cF_\b^j{}_a| = - \xi^\a_j (\g_a)_{\a\b} \l^{ j\b} \ ,
\end{align}
\esubeq
where we have used $\cD_a$ to mean its projection, $\cD_a| = \partial_a + \ri v_a$, when acting 
on a component field.

%%%%%%%%%%%%%%%%%%%%%%%%%%%%%%%%%%%%%%%%%%%%%%%%%%%%%%
%%%%%%%%%%%%%%%%%%%%%%%%%%%%%%%%%%%%%%%%%%%%%%%%%%%%%%

\subsection{The superform formulation for the linear multiplet}

The linear 
multiplet can be described using a four-form gauge 
potential $B = \frac{1}{4!} \rd z^D \wedge \rd z^C \wedge \rd z^B \wedge \rd z^A B_{ABCD}$ 
possessing the gauge transformation
\be \d \cB = \rd \r \ ,
\ee
where the gauge parameter $\r$ is an arbitrary  three-form.\footnote{The construction here 
is a straightforward generalisation of the ones given in \cite{BKN}.}  
The corresponding field strength is
\be H = \rd \cB = \frac{1}{5!} \rd z^E \wedge \rd z^{D} \wedge \rd z^C \wedge \rd z^B \wedge \rd z^A H_{A B C D E} \ ,
\ee
where
\be
H_{A B C D E} = 5 D_{[A} \cB_{BCDE\}} - 10 T_{[A B}{}^{F} \cB_{|F|CDE\}} \ .
\ee
The field strength must satisfy the Bianchi identity
\be 
\rd H = 0  \quad \Longleftrightarrow \quad
D_{[A} H_{BCDEF \}} - \frac{5}{2} T_{[AB}{}^{G} H_{|G|CDEF\}} = 0 \ .
\ee

In order to describe the linear
multiplet we need to impose some covariant constraints on the field strength $H$. 
We choose the constraint
\begin{align}
H_{abc}{}_\a^i{}_\b^j 
= - 2 \ri (\g_{abc})_{\a\b} L^{ij}
\ , \quad L^{ij} = L^{ji} \ , \label{O(2)constrants}
\end{align}
and require all lower mass-dimension components to vanish. 
We can now solve for the remaining components of $H$ in terms of $L^{ij}$. The solution is
\bea
H_{abcd}{}_\a^i
&=& 
- \frac{1}{6} \eps_{abcdef} (\g^{ef})_\a{}^\b D_{\b j} L^{ij}
\ , \\
H_{abcde} &=& - \frac{\ri}{24} \eps_{abcdef} ({\tilde{\g}}^f)^{\a\b} D_\a^k D_\b^l L_{kl} \ ,
\eea
where $L^{ij}$ satisfies the constraint for the 
linear
multiplet
\be D_{\a}^{(i} L^{jk)} = 0 \ .
\ee

We defer the definition of the component fields
and the explicit form of their supersymmetry transformations  to the next subsection.

%%%%%%%%%%%%%%%%%%%%%%%%%%%%%%%%%%%%%%%%%%%%%%%%%%%%%%
%%%%%%%%%%%%%%%%%%%%%%%%%%%%%%%%%%%%%%%%%%%%%%%%%%%%%%

\subsection{Chern-Simons couplings to the linear multiplet}

Unlike in lower dimensions, the linear multiplet in six dimensions permits a deformation with the 
use of a non-abelian vector multiplet. The deformed multiplet we will refer to as the deformed linear multiplet. 
To deform the linear multiplet we now introduce a gauge four-form 
$\cB = \frac{1}{4!} \rd z^D \wedge \rd z^C \wedge \rd z^B \wedge \rd z^A \cB_{ABCD}$ and its five-form 
field strength defined by
\be \cH := \rd \cB + \kappa \, \tr (\cV \wedge \cF \wedge \cF 
+ \frac{\ri}{2} \cV \wedge \cV \wedge \cV \wedge \cF 
- \frac{1}{10} \cV \wedge \cV \wedge \cV \wedge \cV \wedge \cV) \ ,
\ee
where $\cV$ and $\cF$ are the Yang-Mills connection and two-form field strength 
of a non-abelian vector multiplet, respectively. 
Here $\cB$ is understood to be a gauge singlet, $\cD \cB = \rd \cB$. The infinitesimal gauge-transformations are
\bsubeq
\bea
\d \cV &=& - \rd \t \ , \\
\d \cB &=& \rd \r - \kappa \, \tr \Big(\rd \t \wedge (\cV \wedge \cF + \frac{\ri}{2} \cV \wedge \cV \wedge \cV) \Big) \ ,
\eea
\esubeq
where $\t$ and $\r$ generate the gauge transformations of $\cV$ and $\cB$, respectively. The 
field strength $\cH$ satisfies the Bianchi identity
\be \rd \cH = \kappa \, \tr (\cF \wedge \cF \wedge \cF) \ ,
\ee
which is equivalent to
\be 2 D_{[A} \cH_{BCDEF\}}
- 5 T_{[AB}{}^{G} \cH_{|G|CDEF\}}
= 30\,\kappa \, \tr \Big(\cF_{[AB} \cF_{CD} \cF_{EF \}} \Big) \ .
\ee

In order to construct an irreducible multiplet one should constrain the 
components of $\cH$. We can make use of similar constraints as those for the linear multiplet, eq. 
\eqref{O(2)constrants}. We impose the constraint
\begin{align}
\cH_{abc}{}_\a^i{}_\b^j 
= - 2 \ri (\g_{abc})_{\a\b} \cL^{ij}
\ , \quad \cL^{ij} = \cL^{ji}
\end{align}
and require all lower components to vanish. Here $\cL^{ij}$ is a gauge singlet.

Upon imposing these constraints the remaining components of $\cH$ are completely 
determined and are found to be
\bsubeq
\bea
\cH_{abcd}{}_\a^i
&=& 
- \frac{1}{6} \eps_{abcdef} (\g^{ef})_\a{}^\b D_{\b j} \cL^{ij} \non \\
&&+ 
\, \kappa \, \ri\, \eps_{abcdef} (\g^e)_{\a\b} (\g^f)_{\g\d} \tr \Big( \cW^\b_j \cW^{(j \g } \cW^{i) \d } \Big)
\ , ~~~~~\\
\cH_{abcde} &=& \eps_{abcdef} \tilde{H}^f
\ ,
\eea
\esubeq
where
\be D_\a^{(i} \cL^{jk)} = \kappa \, \ri \, \eps_{\a\b\g\d} \tr \Big( \cW^{i\b } \cW^{j\g } \cW^{k\d } \Big)
\label{consisten-deform-L}
\ee
and
\be \tilde{\cH}^a = - \frac{\ri}{24} ({\tilde{\g}}^a)^{\a\b} D_\a^k D_\b^l \cL_{kl}
- \frac{\kappa \, \ri}{2} \tr \Big( X_{kl} (\cW^{k} \g^a \cW^{l}) \Big)
+ \frac{3 \kappa \, \ri}{8} \tr\Big( \cF_{bc} (\cW^{ k} \g^{abc} \cW_k) \Big) \ .~~~
\ee
In deriving the components of $\cH$ the following identity proves useful:
\bea
D_\a^i \Phi_\b^j
&=& - \frac{1}{2} \eps^{ij} D_{[\a}^k \Phi_{\b] k}
- \frac{\kappa}{2} \eps_{\a\b\g\d} \tr\Big( X_k{}^{(i} [ \cW^{j)\g },  \cW^{k \d }] \Big)
+ \ri \partial_{\a\b} \cL^{ij} \non\\
&&- \kappa \, (\g^a)_{\a\b} (\g^b)_{\g\d} \tr\Big(\cF_{ab} \cW^{(i \g } \cW^{j) \d } \Big)
- \frac{\kappa}{4} (\g_{abc})_{\a\b} (\g^a)_{\g\d} \tr \Big( \cF^{bc} \cW^{(i\g } \cW^{j)\d } \Big) \label{deformO(2)cons}
\ ,~~~~~~
\eea
where we have defined
\be \Phi_\a^i = \frac{1}{3} D_{\a j} \cL^{ij} \ .
\ee
It should be noted that the Bianchi identities imply that $\tilde{\cH}^a$ satisfies
\be 
\partial_a \tilde{\cH}^a = \frac{\kappa}{8} \eps^{abcdef} \tr\Big( \cF_{ab} \cF_{cd} \cF_{ef} \Big) \ .
\label{2.366}
\ee

Using the above results one can deduce the supersymmetry transformations. We 
define the independent component fields of $\cL^{ij}$ as follows:
\begin{align}
J^{ij} := \cL^{ij}| \ , \quad \varphi_\a^i := \Phi_\a^i| \ , \quad b_{abcd} := B_{abcd}| \ .
\end{align}
The component projection of the field strength is related to $b_{abcd}$ as follows
\be \cH_{abcde}| = 
5 \partial_{[a} B_{bcde]}
+ \k \, \tr \Big(30 v_{[a} f_{bc} f_{de]} + 30 \ri v_{[a} v_b v_c f_{de]}
+ 12 v_{[a} v_b v_c v_d v_{e]}\Big) \ .
\ee
The supersymmetry transformations of the component fields are found with the help of the 
superform $\cH$ and the identity \eqref{deformO(2)cons}. They are
\bsubeq \label{dO(2)SUSY}
\bea
\d_\xi J^{ij} &=& - 2 \xi^\a_k \varphi_\a^k 
+ \kappa\, \ri \, \xi^\b_k \eps_{\b\g\d\r} \tr\Big( \l^{i\g } \l^{j\d } \l^{k\g } \Big)\ , \\
\d_{\xi} \varphi_\a^i &=& 
- \xi^{\b i} (\g^a)_{\b\a} \tilde{\cH}_a|
+ \kappa \, \xi^{\b i} \eps_{\b\a\g\d} \tr\Big( y_{kl} \l^{k\g } \l^{l\d } \Big)
- 3 \, \kappa \, \xi^{\b i} \eps_{\b\a\g\r} \tr \Big(f_\d{}^\r [\l^{k\g },  \l^\d_k] \Big) \non\\
&&- \frac{\kappa}{2} \xi^\b_j \eps_{\b\a\g\d} \tr\Big( y_k{}^{(j} [ \l^{i)\g },  \l^{k \d }] \Big) + \ri \xi^\b_j  \partial_{\b\a} J^{ji}
- \kappa \, \xi^\b_j (\g^a)_{\b\a} (\g^b)_{\g\d} \tr\Big(f_{ab} \l^{(j\g } \l^{i)\d } \Big) \non\\
&&- \frac{\kappa}{4} \xi^\b_j  (\g_{abc})_{\b\a} (\g^a)_{\g\d} \tr \Big( f^{bc} \l^{(j\g } \l^{i)\d } \Big) 
\ , \\
\d_{\xi} b_{abcd} &=& 
- \frac{1}{2} \eps_{abcdef} \xi^\a_k (\g^{ef})_\a{}^\b \varphi_\b^k
+ \kappa \, \ri \, \eps_{abcdef} \xi^\a_j (\g^e)_{\a\b} (\g^f)_{\g\d} \tr \Big( \l^\b_k \l^{(j\g } \l^{k)\d } \Big) \non\\
&&- 24 \, \k \, \tr \Big( \d_\xi v_{[a} v_{b} f_{cd]} + \frac{\ri}{2} \d_\xi v_{[a} v_{b} v_{c} v_{d]}  \Big) \ .
\eea
\esubeq
The covariant component field strength,
\be 
j^a
:= -2\tilde{\cH}^a| \ ,
\label{2.399}
\ee
transforms as follows:
\bea 
\d_\xi j^a 
&=& -2 (\g^{ab})_\a{}^\b \xi^\a_i \partial_b \varphi_\b^i
+4 \ri \k \xi^\a_i (\g^{[a})_{\a\b} (\g^{b]})_{\g\d} \partial_b \tr\Big( \l^\b_j \l^{(j \g} \l^{i) \d} \Big) \non\\
&&+ \frac{3 \kappa}{2} \eps^{abcdef} \xi^{\a}_i (\g_b)_{\a\b} \tr\Big( \l^{i \b} f_{cd} f_{ef} \Big) \ .
\eea
The supersymmetry transformations for the usual linear multiplet may be obtained by switching off the 
coupling to the vector multiplet, $\kappa =0$.

The component field $j^a$ defined by \eqref{2.399} is normalised  such that 
the bar-projection of \eqref{2.366} in the abelian case 
coincides with \eqref{1.5} upon identifying $\k = 1/(96\p^3)$.

%%%%%%%%%%%%%%%%%%%%%%%%%%%%%%%%%%%%%%%%%%%%%%%%%%%%%%

\section{The anomalous effective action}

Let $\G$ be an effective action for the abelian vector multiplet. 
One may think of $\G$ as the functional obtained by integrating out
 the hypermultiplets coupled to the vector multiplet. 
 
 \subsection{Conventional superspace formulation}
If the vector multiplet is described  
by Mezincescu's prepotential\footnote{See Appendix \ref{AppA} for
a brief review of the known prepotentials for the vector multiplet.} 
\cite{Mezincescu}
$M_{ij}(z)$,
the effective action is a functional of this superfield, 
\be
\Gamma = \Gamma[M_{ij}]~.
\ee
Varying $\G$ leads to the functional derivative
$L^{ij}=L^{ji}$ defined by 
\be
\delta\Gamma = \int \rd^{6|8}z \,\delta M_{ij} L^{ij}~,
\ee
where the integration is performed over the full superspace. 
The Mezincescu prepotential $M_{ij}$ has 
dimension $-2$, and its  gauge transformation is given by eq. \eqref{Mez-gauge},
\be
\delta_\xi M_{ij} = D^k_\alpha \xi^\alpha_{ijk}~,
\label{3.33}
\ee
with the gauge parameter $\xi^\alpha_{ijk}$ being unconstrained.
This gauge transformation means that the theory under consideration is  
a gauge theory with linearly dependent generators, 
following the terminology of the Batalin-Vilkovisky quantisation 
\cite{BV}. Indeed, the gauge parameter in \eqref{3.33} is defined modulo  
arbitrary shifts $ \xi^\alpha_{ijk} \to  \xi^\alpha_{ijk} + \d \xi^\alpha_{ijk}$
of the form 
\bea
\d \xi^\alpha_{ijk} = D_\b^l  \z^{\alpha \b}_{ijkl}~,
\qquad  \z^{\alpha \b} _{ijkl} =  \z^{(\alpha \b)}_{(ijkl)}
\label{3.44}
\eea
such that $\delta_\xi M_{ij} = \delta_{\xi +\d \x}M_{ij} $.
In its turn, the parameter $\z^{\alpha \b} _{ijkl} $ in \eqref{3.44} 
is defined modulo  
arbitrary shifts $ \z^{\alpha \b} _{ijkl} \to  \z^{\alpha \b} _{ijkl} +\d  \z^{\alpha \b} _{ijkl} $,
where
\bea
\d  \z^{\alpha \b} _{ijkl} = D_\g^m  \o^{\alpha \b \g} _{ijkl m} ~, \qquad 
\o^{\alpha \b \g} _{ijkl m} = \o^{( \alpha \b \g )} _{( ijkl m )}~, 
\eea
and so forth. This means that the 6D $\cN=(1,0) $ supersymmetric Yang-Mills 
theory formulated in conventional superspace
is  a gauge theory of infinite degree of reducibility, similar to the Green-Schwarz superstring. 

Under the gauge transformation \eqref{3.33}  the effective action varies as
\be
\delta_\xi\Gamma = \int \rd^{6|8}z \,
\xi^{\alpha}_{ijk} D^{(k}_\alpha L^{ij)}~.
\label{delta-Gamma-an}
\ee
For anomaly-free theories, the effective action is gauge invariant,
which means that $L^{ij}$ obeys the conservation equation \eqref{1.1}.
Therefore $L^{ij}$ is a linear multiplet containing a conserved current. 

In the presence of anomalies, $L^{ij}$ is no longer a linear multiplet. 
Instead it obeys the anomalous conservation equation \eqref{1.2a}. 
In order for the gauge variation 
\be
\delta_\xi\Gamma = \int \rd^{6|8}z \,
\xi^{\alpha}_{ijk} A_\alpha^{ijk}~
\label{delta-Gamma-an2}
\ee
to be invariant under the transformation \eqref{3.44},  the anomaly superfield
$A_\alpha^{ijk}$ must obey the consistency condition \eqref{consistency}. 
Of course, it must also comply with the Wess-Zumino consistency condition
\cite{WZ}. Both conditions are satisfied if the anomaly superfield is
\bea
A^{ijk}_\alpha = \ri\, \k\, \varepsilon_{\alpha\beta\gamma\delta}
\cW^{i\beta} \cW^{j\gamma} \cW^{k\delta}~,
\label{anomaly_}
\eea
for some parameter $\k$.

For anomalous theories, the effective action $\Gamma[M_{ij}]$ 
may be represented  as the sum of two parts, 
\bea
\G = \G_A +\widetilde{\G}~, 
\label{3.99}
\eea
 where $\G_A$ contains all information about the anomaly, 
while $\widetilde{\G}$ is a gauge-invariant functional, 
\be
\delta_\xi \Gamma_A = 
\int \rd^{6|8}z  \, \xi^\alpha_{ijk} A^{ijk}_\alpha~, \qquad
\delta_\xi \widetilde{\Gamma}=0~.
\label{GammaA_}
\ee
Decomposition \eqref{3.99} is not unique.  
The anomalous part of the effective action, $\G_A$, 
may be determined by making the ansatz
\be
\Gamma_A = \int \rd^{6|8}z\, M_{ij}\Lambda^{ij}~, 
\label{GammaA}
\ee
in which $\Lambda^{ij} [\cW^\g_k]$ is a functional of the field strength $\cW^\g_k$
subject to the equation
\be
D^{(k}_\alpha \L^{ij)} = A^{ijk}_\alpha ~.
\label{3.122}
\ee
In Appendix \ref{AppB} we demonstrate that $\L^{ij}$ may be chosen in the form:
\bea
\Lambda^{ij}&=&
\frac{3\ri}8 \frac{\partial^{\alpha\beta}}{\square}
D_{k\alpha} A^{ijk}_\beta
+\frac3{80} \frac1\square \varepsilon^{\alpha\beta\gamma\delta} D_{k\alpha} D_{l\beta} D^{(i}_{\gamma} A^{jkl)}_{\delta} 
-\frac3{160} \frac{\partial^{\alpha\beta}\partial^{\mu\nu}}{\square^2} D_{k\alpha} D_{l\beta} D^{(i}_\mu 
A^{jkl)}_\nu
\non\\&&
+\frac{\ri}{1152}\frac{\partial_{\alpha\alpha'}}{\square^2}
\varepsilon^{\alpha\beta\gamma\delta}
 \varepsilon^{\alpha'\beta'\gamma'\delta'}
 D_{k\beta} D_{l\gamma} D_{m\delta}
 D^{(i}_{\b'} D^j_{\g'} A^{klm)}_{\delta'}
 \non\\&&
-\frac1{64512} \frac1{\square^2} \varepsilon^{\alpha\beta\gamma\delta}
\varepsilon^{\alpha'\beta'\gamma'\delta'}
D_{k\alpha} D_{l\beta} D_{m\gamma} D_{n\delta}
 D^{(i}_{\alpha'} D^j_{\beta'} D^k_{\gamma'} A^{lmn)}_{\delta'}
~.
\label{non-local-solution}
\eea

\subsection{Harmonic superspace formulation}

In the harmonic superspace approach,
the effective action 
for the vector multiplet
is a functional of the analytic prepotential $V^{++} (z,u^\pm_i)$,
\be
\Gamma = \Gamma[V^{++}]~.
\ee
Varying $\G$ 
with respect to $V^{++}$ leads to the functional derivative  $L^{++}$ defined by
\be
\delta\Gamma = \int \rd\zeta^{(-4) } \delta V^{++} L^{++}~,\qquad
D^+_\alpha L^{++} = 0~.
\ee
In particular, for the gauge variation 
$\d_\l V^{++} = - D^{++} \l$, which is the infinitesimal 
form of \eqref{A.122} in the abelian case, we have 
\be
\delta_\lambda \Gamma = \int \rd\zeta^{(-4)}
\delta_\lambda V^{++} L^{++} =
\int \rd\zeta^{(-4)}
\lambda D^{++} L^{++} ~.
\ee
If the theory is anomaly-free, the effective action is gauge invariant, 
$\delta_\lambda \Gamma =0$, and $L^{++}$ obeys the conservation equation
\be
D^{++} L^{++} =0~.
\label{harm-shortness}
\ee
In the central basis, this equation is equivalent to 
\bea
L^{++} (z,u) = L^{ij}(z) u^+_i u^+_j ~.
\eea
The analyticity condition $D^+_\alpha L^{++} = 0$ means that $L^{ij}$
obeys the conservation equation \eqref{1.1}. The conserved current multiplet, 
$L^{ij}$, coincides with the one originating within the conventional superspace formulation described in the previous subsection.

In the presence of anomalies, the conservation equation \eqref{harm-shortness}
is replaced with 
\be
D^{++} L^{++} = A^{(+4)}~,\qquad
D^+_\alpha A^{(+4)} =0~,
\label{harm-anomaly}
\ee 
with  the analytic superfield $A^{(+4)}$ containing all information about  the anomaly. 
The anomaly must obey the Wess-Zumino consistency condition, 
$[\d_{\l_1} ,\d_{\l_2}] \G =0$, which is equivalent to 
\bea
\d_\l A^{(+4)} (\z) =  \int \rd \tilde \zeta^{(-4)} A^{(4,4)} (\z, \tilde \z) \l(\tilde \z) ~,
\qquad A^{(4,4)} (\z, \tilde \z)  = A^{(4,4)} ( \tilde \z , \z)  ~,
\label{3.200}
\eea
for some bi-analytic kernel $ A^{(4,4)} (\z, \tilde \z)$.

%%%%%%%%%%%%%%%%%%%%%%%%%%%%%%%%%%%%%%%%%%%%%%%%%%%%%%

\subsection{Projective superspace formulation}

In the projective superspace approach,
the effective action 
for the vector multiplet
is a functional of the tropical prepotential $V (z,v_i)$,
\be
\Gamma = \Gamma[V]~.
\ee
Varying $\G$ 
with respect to $V$ leads to the functional derivative  $L^{(2)}(z,v)$, 
which is  a weight-2 projective multiplet, 
  defined by
\bea
\d \G &=& \frac{1}{2\pi}  \oint_C (v,\rd v)
\int \rd^6 x \, D^{(-4)}\Big\{\d V  L^{(2)} \Big\}~,\qquad 
D^{(1)}_{\a} L^{(2)} =0~,
\label{proj_var}
\eea
with 
 $C$ a closed integration contour. 
Here we have also introduced the fourth-order operator
\bea
D^{(-4)}:=-\frac{1}{ 96}\ve^{\a\b\g\d}
D^{(-1)}_{\a} D^{(-1)}_{\b}D^{(-1)}_{\g}D^{(-1)}_{\d}~, \qquad
D^{(-1)}_{ \a} := \frac{u_i}{(v,u)} D^i_{ \a}~,
\eea
which involves a constant isospinor $u_i$ constrained by the only condition 
$(v,u) \neq 0$ along the integration contour in \eqref{proj_var}.
The variation  \eqref{proj_var} may be shown to be invariant 
under arbitrary projective transformations 
\be
(u_i\,,\,v_i)~\to~(u_i\,,\, v_i )\,R~,~~~~~~R\,=\,
\left(\begin{array}{cc}a~&0\\ b~&c~\end{array}\right)\,\in\,{\rm GL(2,\mathbb{C})}~,
\ee
and therefore  \eqref{proj_var} is independent of $u_i$. 
It may also be shown that   \eqref{proj_var} is independent of the superspace Grassmann variables.

Choosing $\d V$ in \eqref{proj_var}
to be an infinitesimal gauge variation \eqref{A.29} 
gives
\bea
\d_\l  \G &=& \frac{\ri}{2\pi}  \oint_C (v,\rd v)
\int \rd^6 x \, D^{(-4)}\Big\{(\breve{\l} -\l)  L^{(2)} \Big\}~.
\label{proj_var2}
\eea
If the theory is anomaly-free, the effective action is gauge invariant, 
$\d_\l  \G =0$, for arbitrary weight-0 arctic superfield $\l$. 
It turns out that this condition implies
\bea
L^{(2)} (z,v) = L^{ij}(z) v_i v_j ~.
\label{3.26}
\eea
Then the analyticity condition $D^{(1)}_{\a} L^{(2)} =0$ means that 
$L^{ij}$ obeys the constraint \eqref{1.1}.
The conserved current multiplet, 
$L^{ij}$, coincides with those originating within the conventional 
and harmonic superspace formulations described in the previous subsections.
Eq. \eqref{3.26} tells us that associated with the conserved current multiplet 
$L^{ij}$, eq. \eqref{1.1}, 
 is the  holomorphic tensor field $L^{(2)}$ over ${\mathbb C}P^1$.

If the theory is anomalous, the gauge variation \eqref{proj_var2} does not vanish. 
As a consequence, the projective multiplet $L^{(2)} $ is no longer a linear multiplet. 

\section{An alternative description of the anomalous current multiplet}

In section \ref{section2} we have constructed the consistent deformation 
of the 6D $\cN=(1,0)$ linear multiplet given by eq. 
\eqref{consisten-deform-L}.
Here an alternative form for the anomalous current multiplet will be derived in 
the abelian case. 
We will use some harmonic superspace relations described in 
subsection \ref{subsectionA.1}.

We  associate with the anomalous current multiplet $L^{ij}$, 
eq. \eqref{1.4}, 
 the following harmonic superfield:
\bea
{ L}^{++} = u^+_i u^+_j { L}^{ij}\,, \qquad D^{++} L^{++}=0~.
\label{3.2}
\eea
Then eq. \eqref{1.4} is equivalent to 
\bea
D^+_\alpha { L}^{++}
=  \kappa \, \ri \,\varepsilon_{\alpha\beta\gamma\delta} \cW^{+\beta} \cW^{+\gamma} \cW^{+\delta}\,,
\label{J-A}
\eea
with the superfield $\cW^{+ \alpha} $ being defined by \eqref{Harm-proj}.

In the anomaly-free case, the current multiplet ${ L}^{++} = u^+_i u^+_j { L}^{ij}$
is analytic and holomorphic on ${\mathbb C}P^1 $, 
\bea
D^+_\a L^{++} =0~, \qquad D^{++} L^{++} =0~.
\eea
Eq. \eqref{J-A} tells us that the anomalous current multiplet is no longer analytic. 

As a first step,  we represent 
\bea
\ri \,\varepsilon_{\alpha\beta\gamma\delta} \cW^{+\beta} \cW^{+\gamma} \cW^{+\delta}
= D^+_\a F^{++}~, 
\label{3.4}
\eea
for some superfield $F^{++}(z, u^\pm)$ defined up to an arbitrary shift 
of the form
\bea
F^{++} ~\to ~F^{++} + H^{++} ~, \qquad D^+_\a H^{++} =0~.
\eea
A particular solution of  \eqref{3.4} is 
\bea
{F}^{++} =-\frac\ri2 V_{\alpha\beta} \cW^{+\alpha} \cW^{+\beta}
-  \frac{\ri}{64} \varepsilon^{\alpha\beta\gamma\delta} 
 V_{\alpha\beta} V_{\gamma\delta} D^+ \cW^+
 ~,
\label{4.77}
\eea
where $V_{\alpha\beta}$ is the vector superfield connection
defined in (\ref{Vab}). 
In checking (\ref{3.4}) the following properties of $V_{\a\b}$ may be useful
\be
D^+_\alpha V_{\beta\gamma} = -2
\varepsilon_{\alpha\beta\gamma\delta} \cW^{+\delta}~,\qquad
D^{++} V_{\alpha\beta} =
{\pa}_{\a\b} V^{++}~.
\ee
It is seen that $F^{++}$  is neither analytic nor gauge invariant. 
However, 
$D^{++} F^{++} $ proves to be  analytic, 
\bea
D^{++} {F}^{++}  
&=&-\frac{ \ri}{2}
 G^{++\alpha\beta}  {\pa}_{\a\b} V^{++}
~,
\label{AA}
\eea
where we have defined 
\bea
G^{++\alpha\beta} = \cW^{+\alpha}\cW^{+\beta} +\frac1{16}\varepsilon^{\alpha\beta\gamma\delta}V_{\gamma\delta}D^+ \cW^+~,\qquad
D^+_\gamma G^{++\alpha\beta} &=& 0~.
\eea

Our second step is to  introduce
\bea
{\mathbb L}^{++}= L^{++} - \kappa\,F^{++} ~.
\eea
It follows from \eqref{J-A} and \eqref{3.4}
that ${\mathbb L}^{++}$ is analytic, 
\bea
D^+_\a {\mathbb L}^{++} =0~.
\eea
However, unlike $L^{++}$,  the superfield ${\mathbb L}^{++}$
 is no longer holomorphic on ${\mathbb C}P^1 $, 
\bea
D^{++} {\mathbb L}^{++} = {\mathbb A}^{(+4)}~, \qquad
D^+_\a {\mathbb A}^{(+4)} =0~.
\eea
The anomaly is now encoded in the analytic superfield ${\mathbb A}^{(+4)} $.
 It is defined modulo shifts
\bea
{\mathbb A}^{(+4)} ~\to ~ {\mathbb A}^{(+4)} - \kappa\, D^{++}H^{++}~,\qquad 
D^+_\a H^{++}=0~,
\label{4.133}
\eea
where the analytic superfield $H^{++}$ is a {\it local} composite of the gauge prepotential.

For the choice of $F^{++}$ given above, eq. \eqref{4.77}, ${\mathbb A}^{(+4)} $
is 
\bea
{\mathbb A}^{(+4)} = \frac{ \ri}{2} \k 
 G^{++\alpha\beta} 
{\pa}_{\a\b} V^{++}
~.
\eea
It is an interesting problem to understand whether the functional freedom 
\eqref{4.133} allows one to construct an analytic superfield
${A}^{(+4)} = {\mathbb A}^{(+4)} - \kappa\, D^{++}H^{++}$
obeying the Wess-Zumino consistency condition 
\eqref{3.200}.

%%%%%%%%%%%%%%%%%%%%%%%%%%%%%%%%%%%%%%%%%%%%%%%%%%%%%%%

\section{Concluding comments}

In this paper we have presented the  consistent deformation,
eq. \eqref {consisten-deform-L},  of the 
6D $\cN=(1,0)$ linear multiplet which describes chiral anomalies. It is
\be 
  D_\a^{(i} \cL^{jk)} = \kappa \, \ri \, \eps_{\a\b\g\d}\, \tr \Big( \cW^{i\b } \cW^{j\g } \cW^{k\d } \Big)~.
\label{5.1}
\ee
Its consistency 
is guaranteed by 
the superform formulation 
for the deformed linear multiplet developed in section \ref{section2}.
Equation \eqref{5.1} is superconformal
assuming $\cL^{ij} $ to be a primary superfield of dimension 4.

The consistent Chern-Simons coupling of the linear multiplet to a 
vector multiplet, eq. \eqref{5.1}, is a characteristic feature of 6D $\cN=(1,0)$
supersymmetry.
Such a deformation was not possible in the cases 
of 4D $\cN=2$ and 5D $\cN=1$ supersymmetry. 
Equation \eqref{5.1}  is analogous
to the constraint describing a deformed 4D $\cN=1 $ linear multiplet
$\cL = \bar \cL$, which is
\bea
\bar D^2 \cL = 2\k\, {\rm tr} \big( \cW^\a \cW_\a \big)~, \qquad
 D^2 \cL = 2\k\, {\rm tr} \big( \bar \cW_\ad \bar \cW^\ad \big)~, 
\eea
with $\cW_\a$ the covariantly chiral field strength of a non-abelian vector multiplet, 
see \cite{BGG} for a review of the Chern-Simons couplings to the 4D $\cN=1 $
linear multiplet.

In the abelian case, we have computed the nonlocal effective action $\G_A$, 
which is  given by the relations \eqref{GammaA} and \eqref{non-local-solution}
and which generates the gauge anomaly \eqref{anomaly_}.
The effective action $\G_A$ is constructed as a functional of the 
Mezincescu prepotential, which corresponds to the formulation 
of the 6D $\cN=(1,0) $ vector multiplet in conventional superspace 
\cite{HSierraT,Koller}. It is known that such a formulation is not suitable
(unlike, e.g., the harmonic superspace approach) 
to do quantum calculations in general supersymmetric Yang-Mills theories
with eight supercharges in diverse dimensions. There are many reasons for that, 
and the most prominent ones are the following. 
Firstly, the conventional superspace 
 approach does not offer means to describe off-shell hypermultiplets
in complex representations of the gauge group (see \cite{GIOS} for a detailed discussion). Secondly, the Yang-Mills multiplet in this approach is a 
nontrivial gauge theory with linearly dependent generators of infinite degree of reducibility. As discussed in detail in \cite{BKO}, the Batalin-Vilkovisky quantisation 
of the theory has never been used to derive a consistent
superfield effective action.\footnote{Even the
Batalin-Vilkovisky quantisation scheme \cite{BV} is literally applicable to finitely reducible gauge theories only.} 
Both problems simply do not occur with the harmonic superspace
and the projective superspace approaches. 

We computed the effective action  \eqref{GammaA}, \eqref{non-local-solution}
by integrating the gauge anomaly \eqref{anomaly_}.
However, we did not compute the anomaly by doing supergraph calculations. 
Once the structure of the anomalous current multiplet is established, 
it suffices to make 
use of the known non-supersymmetric results
\cite{Frampton,Leutwyler}. This is exactly what was done in this paper. 
It is of interest to compute the gauge anomaly by doing direct supergraph 
calculations in 6D $\cN=(1,0)$ harmonic superspace, however 
the existing literature \cite{ISZ,BP} does not offer any insight. 
We hope to report on such calculations elsewhere.

%%%%%%%%%%%%%%%%%%%%%%%%%%%%%%%%%%%%%%%%%%%%%%%%%%%%%%
%$~$\\
\noindent
{\bf Acknowledgements:}\\ 
JN is grateful to Daniel Butter for useful discussions.
This work is supported in part by the Australian Research Council, project No.
 DP140103925. 
 JN acknowledges support from GIF -- the German-Israeli Foundation for Scientific Research and Development.

%%%%%%%%%%%%%%%%%%%%%%%%%%%%%%%%%%%%%%%%%%%%%%%%%%%%%%
%%%%%%%%%%%%%%%%%%%%%%%%%%%%%%%%%%%%%%%%%%%%%%%%%%%%%%

\appendix

%%%%%%%%%%%%%%%%%%%%%%%%%%%%%%%%%%%%%%%%%%%%%%%%%%%%%%
%%%%%%%%%%%%%%%%%%%%%%%%%%%%%%%%%%%%%%%%%%%%%%%%%%%%%%

%
\section{Prepotentials for the Yang-Mills multiplet}
\label{AppA}

In the case of supersymmetry with eight supercharges
in diverse dimensions, $3\leq d \leq 6$, 
there exist three different prepotential 
formulations for the Yang-Mills multiplet, which make use of the following 
multiplets: (i) the Mezincescu prepotential \cite{Mezincescu}; 
(ii) the analytic prepotential \cite{GIKOS}; 
and (iii) the tropical prepotential \cite{LR2}.
The Mezincescu prepotential can be obtained from the analytic one 
as described in section 7.2.4 of \cite{GIOS}. 
It can also be read off from the tropical prepotential in accordance
with \cite{BK11}. In its turn, the tropical prepotential can be obtained
from the analytic one by getting rid of an infinite tail of 
superfluous gauge degrees of freedom \cite{K98}.
In spite of these relationships, the three distinct prepotentials are useful for different applications.

\subsection{Analytic prepotential}\label{subsectionA.1}

Supersymmetric Yang-Mills theory in six-dimensional 
$\cN=(1,0)$ harmonic superspace was formulated in \cite{HSWest,Zupnik86}. 
Here we briefly review this formulation
following the harmonic superspace notation of \cite{GIOS}.

Let $u^+_i$ and $u^-_i$ be  standard SU(2) harmonic variables, 
$ \big({u_i}^-\, ,{u_i}^+\big) \in \rm SU(2)$, 
\bea
\overline{u^{+i}} = u^-_i~,
\qquad u^{+i}u_i^- = 1 \;,
\eea
with $ u^+_i = \ve_{ij}u^{+j}$.
Let $D^{++}$, $D^{--}$ and  $D^0$ be the associated 
harmonic derivatives defined as in \cite{GIOS}.
Using the harmonics we introduce U(1) projections of the gauge-covariant spinor derivatives
\be
{\cal D}^\pm_\alpha = u^\pm_i {\cal D}^i_\alpha
= D^\pm_\alpha + \ri \cV^{\pm}_\alpha~,\qquad
\cV^\pm_\alpha = u^\pm_i \cV^i_\alpha~.
\label{cov-D}
\ee
In accordance with (\ref{gauged-algebra}), the operators (\ref{cov-D}) obey the following (anti)commutation relations%
\begin{subequations}
\bea
\{ {\cal D}^+_\alpha , {\cal D}^+_\beta  \} &=&0~,\label{B5a}\\
\{  {\cal D}^+_\alpha , {\cal D}^-_\beta \} &=& 2\ri 
(\gamma^a)_{\alpha\beta} {\cal D}_a~,\label{B5b}\\
{}[{\cal D}_a , {\cal D}^\pm_\alpha] &=& \ri (\gamma_a)_{\alpha\beta} \cW^{\pm \beta}~,\label{B5c}\\
{}[{\cal D}_a, {\cal D}_b] &=& \ri \cF_{ab}~,
\label{B5d} 
\eea
\end{subequations}
where $\cW^{\pm\alpha}$ are the irreducible U(1) components  of the 
field strength $\cW^{i\alpha}$, 
\bea
\cW^{\pm \alpha} = u^\pm_i \cW^{i\alpha}\,.
\label{Harm-proj}
\eea

In the harmonic superspace setting, it is useful to combine the superspace 
gauge-covariant derivatives with the harmonic ones,
\be
{\cal D}_{\hat A} = ({\cal D}_a , {\cal D}^\pm_\alpha,
{\cal D}^{++},{\cal D}^{--}, {\cal D}^0)
:=({\cal D}_a , {\cal D}^\pm_\alpha,D^{++},D^{--}, D^0)
= {D}_{\hat A} +\ri \cV_{\hat A}
~.
\label{tau-frame}
\ee
The gauge transformation of $\cD_{\hat A} $
is analogous to (\ref{2.2}), 
\be
{\cal D}_{\hat A} \longrightarrow
{\cal D}_{\hat A}  = \re^{\ri \t}{\cal D}_{\hat A} \re^{-\ri \t}~.
\ee
Since the gauge superfield parameter $\tau$ is harmonic independent, the harmonic derivatives 
$(D^{\pm\pm},D^0)$ are gauge covariant.

The equation (\ref{B5a}) is the integrability condition for 
covariantly analytic superfields to exist.
This equation can be solved in terms of a bridge superfield $b = b(z,u)$
 defined by the rule
\be
{\cal D}^+_\alpha = \re^{-\ri b } D^+_\alpha \re^{\ri b}~.
\ee
The introduction of the bridge superfield leads to a new gauge freedom, in addition to the $\tau$-gauge transformations (\ref{2.2}). The complete gauge transformation 
law of $b$ is
\be
\re^{\ri b'} = \re^{\ri\lambda} \re^{\ri b}\re^{-\ri\tau}~,
\ee
where $\lambda$ is a  U(1) neutral analytic superfield, 
$D^+_\alpha\lambda =0$.

The representation (\ref{tau-frame}) for the gauge-covariant derivatives  
is called the $\tau$-frame. The bridge superfield  allows one 
to introduce a new representation for the gauge-covariant  derivatives, which is 
defined by 
\be
{\cal D}_{\hat A} \longrightarrow
\nabla_{\hat A} = \re^{\ri b}{\cal D}_{\hat A} \re^{-\ri b}
= {D}_{\hat A} +\ri V_{\hat A}
\ee
and is called the $\lambda$-frame.
In this frame, the derivative $\nabla^+_\alpha$ is short, 
$\nabla^+_\alpha = D^+_\alpha$, and hence $V_\a^+=0$.
However,  two of the three harmonic derivatives acquire gauge connections:
\be
\nabla^{++} = D^{++} + \ri \,V^{++}~,\qquad
\nabla^{--} = D^{--} + \ri V^{--}~.
\ee
As follows from the commutation relation $[\nabla^+_\alpha , \nabla^{++}]=0$, the gauge connection $V^{++}$ is analytic,
\be
D^+_\alpha V^{++} = 0~.
\ee
The connection
 $V^{--}$ 
 can be expressed via $V^{++}$ as a unique solution of 
 the zero-curvature 
condition
\be
[\nabla^{++}, \nabla^{--}] = D^0 \quad
\Longleftrightarrow \quad
D^{++} V^{--} - D^{--} V^{++} + \ri [V^{++}, V^{--}] =0~.
\label{zero-curv}
\ee
The explicit expression for $V^{--}$ in terms of $V^{++}$ was originally found 
by Zupnik \cite{Zupnik:1987vm}. 
In the $\lambda$-frame, no $\t$-gauge freedom remains. 
Under the $\lambda$-gauge group,
the connections $V^{++}$ and $V^{--}$ transform as
\bea
V'^{\pm\pm}=\re^{{\rm i}\lambda}V^{\pm\pm}\re^{-{\rm i}\lambda}
-\ri \,\re^{{\rm i}\lambda}D^{\pm\pm}\re^{-{\rm i}\lambda}~.
\label{A.122}
\eea
The $\l$-frame counterparts of the field strengths $\cW^{\pm\alpha}$
will be denoted $W^{\pm\alpha}$. In the abelian case, there is no 
difference between $\cW^{\pm\alpha}$ and $W^{\pm\alpha}$.

The $\l$-frame counterparts of
the (anti-)commutation relations (\ref{B5b}) and \eqref{B5c}, 
in conjunction with the identity $[\nabla^{--}, \nabla^+_\a] = \nabla^-_\a$,
allow one to express the gauge connections 
 $V^-_\a$ and $V_a$ and the field strength  $W^{+\alpha}$ 
in terms  $V^{--}$. The explicit expressions for the connections are 
\begin{subequations}
\bea
V^-_\alpha &=& - D^+_\alpha V^{--}~, \\
\qquad
V_a &=& \frac\ri8 (\tilde\gamma_a)^{\alpha\beta}
D^+_\alpha D^+_\beta V^{--} \quad 
\Longleftrightarrow \quad
V_{\a\b} = (\gamma^a)_{\alpha\beta} V_a = 
\frac\ri2 D^+_\alpha D^+_\beta V^{--}~.~~~~~~~
\label{Vab}
\eea
\end{subequations}
The expression for the  field strength  is
\be
W^{+\alpha} = \frac\ri{24} \varepsilon^{\alpha\beta\gamma\delta}
D^+_\beta D^+_\gamma D^+_\delta V^{--}~.
\ee
As mentioned above, $V^{--}$ is uniquely expressed 
in terms of the analytic connection $V^{++}$.
Thus, the superfield $V^{++}$ is a single  prepotential in terms of which all the connections are determined, in complete analogy with the 
4D case \cite{GIKOS}. 
This prepotential is analytic, but otherwise unconstrained. 

%%%%%%%%%%%%%%%%%%%%%%%%%%%%%%%%

\subsection{Mezincescu's prepotential}

The Mezincescu prepotential was used in \cite{HSierraT,Koller} 
to describe the 6D $\cN=(1,0)$ vector multiplet in Minkowski superspace.
 In this subsection 
we recall how the Mezincescu prepotential is obtained from the analytic one 
following the discussion in  section 7.2.4 of \cite{GIOS}. 
Only the abelian vector multiplet is considered here. 

In the harmonic superspace approach,  
the gauge prepotential $V^{++}$ and the gauge 
parameter $\lambda$ are analytic superfields, 
$D^+_\alpha V^{++} = 0$ and $D^+_\alpha \l = 0$.
The analyticity constraint on $V^{++}$  is  solved by 
\be
V^{++} = (D^+)^4 M^{--}~,
\label{A.15}
\ee 
where 
\be
(D^+)^4 = -\frac1{96} \varepsilon^{\alpha\beta\gamma\delta}
D^+_\alpha D^+_\beta D^+_\gamma D^+_\delta
\ee
is the analytic projection operator, and 
$M^{--} (z, u) $ is an unconstrained superfield.
Similarly, the analyticity constraint on $\l$ is  solved by 
\bea
\lambda = (D^+)^4 \rho^{(-4)}~,
\eea
where  $\rho^{(-4)} (z,u) $ is an unconstrained superfield.  
The original $\lambda$-transformation of $V^{++}$ turns into 
the following gauge transformation of $M^{--}$: 
\be
\delta_\lambda M^{--} = -D^{++} \rho^{(-4)}~.
\label{lambda-M}
\ee
In addition, it follows from \eqref{A.15} that $M^{--}$ possesses 
a new gauge freedom 
that  leaves $V^{++}$ invariant and acts on $M^{--}$ as follows:
\be
\delta_\xi M^{--} = D^+_\alpha \xi^{(-3)\alpha}~.
\ee
Here $\xi^{(-3)\alpha}(z,u) $ is an unconstrained gauge parameter.

The superfields $M^{--} (z,u)$, $\rho^{(-4)}(z,u) $ and $\xi^{(-3)\alpha}(z,u)$ 
are smooth scalar fields on the group manifold SU(2) of definite U(1) charges
or, equivalently, smooth tensor fields on the two-sphere $S^2 ={\rm SU(2) /U(1)}$.  
Therefore  these superfields 
are given  by convergent Fourier 
series in the harmonic variables, 
\begin{subequations}
\bea
M^{--}(z,u)&=&M^{ij}(z)u^-_i u^-_j + 
M^{ijkl}(z)u^+_{(i} u^-_j u^-_k u^-_{l)} +\ldots~,\label{M-series}\\
\rho^{(-4)}(z,u)&=& \rho^{ijkl}(z)u^-_i u^-_j u^-_k u^-_l + \ldots
~,\label{rho-series}\\
\xi^{(-3)\alpha}(z,u)&=& \frac43\xi^{ijk\,\alpha}(z) u^-_i u^-_j u^-_k+\ldots~,
\label{xi-series}
\eea
\end{subequations}
where the numerical coefficient  in the last relation 
is introduced for later convenience.
Comparing the series (\ref{M-series}) and (\ref{rho-series}), one can see that
the gauge freedom (\ref{lambda-M}) allows one 
 to gauge away all Fourier components of $M^{--}$ in (\ref{M-series}) 
 except for the lowest one. 
 In other words, one can impose  a supersymmetric gauge 
\be
M^{--} (z,u)= M^{ij} (z)u^-_i u^-_j~,\qquad
M^{ij}=M^{(ij)}~.
\label{M-gauge}
\ee
The remaining superfield $M^{ij}$ is exactly Mezincescu's prepotential
\cite{Mezincescu}. 

The gauge condition  (\ref{M-gauge})  completely fixes the $\r$-gauge freedom. 
However, there remains a residual $\x$-invariance
generated solely by 
the spinor gauge parameter $\xi^{ijk\,\alpha}=\xi^{(ijk)\alpha}$ in the 
series (\ref{xi-series}). It acts on  Mezincescu's prepotential 
by the rule
\be
\delta_\xi M_{ij} =  D^k_{\alpha} \xi_{ijk}^\alpha~,
\label{Mez-gauge}
\ee
which is exactly the gauge transformation derived in 
 \cite{HSierraT,Koller}. 
In order to preserve the gauge condition   (\ref{M-gauge}), this $\x$-transformation 
has to be accompanied by a special $\r$-transformation 
\bea
\rho^{(-4)}(z,u)&=& 
 \frac13 D^{i}_\alpha
\xi^{jkl \,\alpha}(z)
u^-_i u^-_j u^-_k u^-_l ~.
\eea

\subsection{Tropical prepotential}

Here we recall the definition of the tropical prepotential \cite{LR2}
which is used to describe the Yang-Mills multiplet within the projective superspace
approach \cite{LR1,LR2}. We follow the modern presentation of this approach given, 
e.g., in \cite{K-lectures}. 

In the projective superspace setting, one does not work with the harmonics
used in subsection \ref{subsectionA.1}.
Instead, one deals with homogeneous coordinates  
$ v^i \in {\mathbb C}^2 \setminus \{0\}$ 
for ${\mathbb C}P^1$.  
We recall that  ${\mathbb C}P^1$ is obtained from ${\mathbb C}^2 \setminus \{0\}$ 
by factorisation with respect to the equivalence relation  $ v^i \sim c\,v^i$,
with $c\in {\mathbb C}^*$.  
Supersymmetric field theories are described in 
terms of the so-called weight-$n$ projective multiplets $Q^{(n)} (z,v) $.  
By definition, such a superfield 
is defined by the following conditions: \\
(i) $Q^{(n)} (z,v) $
is  {\it holomorphic } over an {\it open domain} of $ {\mathbb C}P^1$,
\bea
\frac{\pa}{\pa {\bar v}_i} \, Q^{(n)} =0~.
\eea
(ii) it is a homogeneous function of $v^i$ of degree $n$, 
\bea
Q^{(n)} (z,c\, v )= c^{n}  \, Q^{(n)}(z,v)~,
\quad c \in {\mathbb C}^* ~.
 \eea
(iii) it obeys the analyticity condition 
\bea
D^{(1)}_\a Q^{(n)} =0~, \qquad D^{(1)}_\a = v_i D^i_\a~.
\eea

Introduce two special points in $N, S\in {\mathbb C}P^1$: the north pole $N$ 
with homogeneous coordinates $v^i \propto (0,1)$, 
and  the south pole $S$ labeled by 
$v^i \propto (1,0)$. Associated with these points are two open domains, 
the north chart ${\mathbb C}P^1 \setminus \{ N \}$ and the south 
chart ${\mathbb C}P^1 \setminus \{S \}$, which cover ${\mathbb C}P^1 $.  
In the north chart, we can introduce a complex 
(inhomogeneous) coordinate $\z$  as 
\bea 
 v^i = v^{\1} \,(1, \z) ~,\qquad \z:=\frac{v^{\2}}{v^{\1}} ~,\qquad\quad 
{ i=\1 ,\2}~.
\label{Zeta}
\eea

The tropical multiplet $V(z,v)$ is a weight-0 projective multiplet holomorphic
on ${\mathbb C}P^1 \setminus \{ N \cup S\}$. It is also constrained to be real under 
the so-called smile conjugation, see  \cite{K-lectures} for more details.
It is given by a Laurent series
\bea
 V(z,v) =  V(z,\z ) 
=\sum_{n=-\infty}^{\infty}  V_n (z) \z^n~, \qquad 
V_n^\dagger = (-1)^n  V_{-n}~.
\eea
The gauge transformation law of the tropical prepotential is 
\bea
{\rm e}^{ V' } =  {\rm e}^{{\rm i}\breve{ \l} }  {\rm e}^{ V }  
{\rm e}^{ -{\rm i}  \l } 
~,
\label{A.29} 
\eea
where the gauge parameter $ \l(z, \z) $ is a  weight-zero arctic multiplet
\bea
D_\a^{(1) } { \l} = 0~, 
\qquad 
{ \l}(z, \z) =  \sum_{n=0}^{\infty} { \l}_n (z)\z^n~,
\eea
and its smile conjugated antarctic multiplet, 
\bea
\breve{\l}{}(z,\z) =  \sum_{k=0}^{\infty} (-1)^k{ \l}_k^\dagger (z) \frac{1}{\z^k}~.
\eea
By definition, a weight-$n$ arctic multiplet is holomorphic 
on ${\mathbb C}P^1 \setminus \{ N\}$. 

Modulo purely gauge degrees of freedom, the gauge-covariant derivatives
can be expressed in terms of the tropical prepotential. 
This is explained in detail in the cases of 3D $\cN=4$
and 5D $\cN=1$ vector multiplets coupled to conformal supergravity 
in \cite{KT-M_2014} and \cite{BKNT-M14} respectively. 
The 6D $\cN=(1,0)$ case can be treated similarly. 

Following  \cite{BK11}, 
the Mezincescu prepotential is introduced by the  rule
\begin{align}
M_{ij} (z) = \frac{1}{2\pi} \oint_C 
(v,\rd v)
\, v_i v_j\, U^{(-4)} (z,v)~, 
\qquad  (v,\rd v) =  v^k {\rm d}v_k ~,
\end{align}
where 
$U^{(-4)} (z,v)$ is related to the tropical prepotential as follows:
\bea
V(z,v) = D^{(4)} U^{(-4)} (z,v)~, 
\qquad 
D^{(4)} = -\frac1{96} \varepsilon^{\alpha\beta\gamma\delta}
D^{(1)}_\alpha D^{(1)}_\beta D^{(1)}_\gamma D^{(1)}_\delta~.
\eea

\section{Derivation of the anomalous effective action}
\label{AppB}

To find the functional generating the anomalous effective action
(\ref{GammaA}) it suffices to find a particular solution of the equation 
(\ref{3.122}) which we denote by $\Lambda^{ij}$. 
In this Appendix we will demonstrate that a particular solution of this equation can be represented in the form (\ref{non-local-solution}). 

Given the superfields $\Lambda^{ij}$ and $A^{ijk}_\alpha$ it is convenient to deal with their harmonic projections
\be
\Lambda^{++} = u^+_i u^+_j \L^{ij}~,\qquad
A^{(+3)}_\alpha = u^+_i u^+_j u^+_k A^{ijk}_\alpha~.
\label{B1}
\ee
Then the equation (\ref{3.122}) is equivalent to
\be
D^+_\alpha \Lambda^{++} = A^{(+3)}_\alpha~,
\label{L-eq}
\ee
where $D^+_\alpha = u^+_i D^i_\alpha$.
Note that, by construction, the superfield $\Lambda^{++}$ obeys
\be
D^{++} \Lambda^{++} =0~.
\label{D++L}
\ee
We also point out that $A^{(+3)}_\alpha$ satisfies
\be
D^+_{(\alpha} A^{(+3)}_{\beta)}=0 \quad \Longrightarrow \quad
(D^+)^4 A^{(+3)}_\alpha =0~,
\ee
as a consequence of (\ref{consistency}).

We look for a solution of the equation (\ref{L-eq}) in the form of the sum of the following terms
\be
\Lambda^{++} = \sum_{i=1}^9 c_i \Lambda_i^{++}~,
\label{L-sum}
\ee
where $c_i$ are some coefficients and 
\begin{subequations}\label{B6}
\bea
\Lambda^{++}_1&=& \frac\ri2\frac{\partial^{\alpha\beta}}{\square}
 D^-_\alpha A^{+++}_\beta~,
 \\
\Lambda^{++}_2 &=& \ri \frac{\partial^{\alpha\beta}}{\square} D^{--}  D^+_\alpha A^{+++}_\beta~,
\\
\Lambda^{++}_3 &=& \frac1{\square} \varepsilon^{\alpha\beta\gamma\delta} D^-_\alpha D^-_\beta D^+_\gamma A^{+++}_\delta~,
\\
\Lambda^{++}_4 &=& \frac1\square D^{--} \varepsilon^{\alpha\beta\gamma\delta} D^-_\alpha D^+_\beta D^+_\gamma A^{+++}_\delta~,
\\
\Lambda^{++}_5 &=& \frac1\square D^{--} D^{--} \varepsilon^{\alpha\beta\gamma\delta} D^+_\alpha D^+_\beta D^+_\gamma A^{+++}_\delta~,
\\
\Lambda^{++}_6 &=& \frac{\partial^{\alpha\beta}\partial^{\mu\nu}}{\square^2} 
 D^-_\alpha D^-_\beta 
  D^+_\mu A^{+++}_\nu~,
  \\
\Lambda^{++}_7 &=&\ri \frac{\partial_{\alpha\alpha'}}{\square^2}
 \varepsilon^{\alpha\beta\gamma\delta}
 D^-_\beta D^-_\gamma D^-_\delta
\varepsilon^{\alpha'\beta'\gamma'\delta'}
 D^+_{\beta'} D^+_{\gamma'}
  A^{+++}_{\delta'}~,
  \\
\Lambda^{++}_8 &=&\ri \frac{\partial^{\mu\nu}}{\square^2} D^{--} 
  D^-_\mu D^-_\nu \varepsilon^{\alpha\beta\gamma\delta} D^+_\alpha D^+_\beta D^+_\gamma A^{+++}_\delta~,
 \\
\Lambda^{++}_9 &=& \frac1{\square^2} \varepsilon^{\alpha\beta\gamma\delta}(D^-)^4 D^+_\alpha D^+_\beta D^+_\gamma A^{+++}_\delta~.
\eea
\end{subequations}
The equation (\ref{D++L}) is satisfied on the condition that 
\begin{subequations}\label{126}
\bea
c_1 +8c_2 +8c_3 +16 c_6 &=&0~,
\\
c_3 +2c_4 +\frac13 c_6 -6c_7 &=&0~,
\\
c_4 +10 c_5 -8 c_8 &=&0~,\\
3c_7+8c_8 -\frac12 c_9 &=&0~.
\eea
\end{subequations}
Imposing the equation (\ref{L-eq}) we find the following constraints
 for the coefficients $c_i$:
 \begin{subequations}\label{128}
\bea
c_1 &=& 1~,\non\\
c_1 +4c_2 +16c_6 &=&0~,
\\
c_1 - 16c_3 &=&0~,
\\
c_2 - 6c_4 &=&0~,
\\
\frac23 c_3 + c_4 -12 c_7 &=&0~,
\\
c_4 +8 c_5 -16 c_8 &=&0~,
\\
c_6 + 18 c_7 &=&0~,
\\
4 c_8 +3 c_7 -c_9 &=&0~.
\eea
\end{subequations}
The solution of the system of equations (\ref{126}) and (\ref{128}) reads
\bea
&&c_1= 1~,\quad
c_2=-\frac1{8}~,\quad
c_3= \frac1{16}~,\quad
c_4=-\frac1{48}~,\quad
c_5= \frac1{576}~,\non\\&&
c_6=-\frac1{32}~,\quad
c_7= \frac1{576}~,\quad
c_8= -\frac1{2304}~,\quad
c_9= \frac1{288}~.
\eea

Note that different terms in 
\eqref{L-sum}
depend on different harmonic monomials. Nevertheless, the equation (\ref{D++L}) guarantees that the full expression (\ref{L-sum}) is quadratic in harmonics in agreement with (\ref{B1}). Therefore, we can restore $\Lambda^{ij}$ from $\Lambda^{++}$ by the rule
\be
\Lambda^{ij} = 3 \int \rd u \, u^{-i} u^{-j} \Lambda^{++}~. 
\ee
The harmonic integral is computed according to the formula 
\cite{GIOS1}
\be
\int \rd u \, u^{+i_1} \ldots u^{+i_n}
 u^-_{j_1} \ldots u^-_{j_n}
 =\frac1{n+1} \delta^{i_1}_{( j_1}
  \ldots \delta^{i_n}_{j_n)}~.
\ee
Applying this rule to all terms in the sum (\ref{L-sum}) we find
\bea
\Lambda^{ij} &=&3\sum_{k=1}^9 c_k \int \rd u\, u^{-i} u^{-j} \Lambda^{++}_k\non\\  
&=&\frac{3\ri}8 \frac{\partial^{\alpha\beta}}{\square}
D_{k\alpha} A^{ijk}_\beta
+\frac3{80} \frac1\square \varepsilon^{\alpha\beta\gamma\delta} D_{k\alpha} D_{l\beta} D^{(i}_{\gamma} A^{jkl)}_{\delta} 
-\frac3{160} \frac{\partial^{\alpha\beta}\partial^{\mu\nu}}{\square^2} D_{k\alpha} D_{l\beta} D^{(i}_\mu 
A^{jkl)}_\nu
\non\\&&
+\frac{\ri}{1152}\frac{\partial_{\alpha\alpha'}}{\square^2}
\varepsilon^{\alpha\beta\gamma\delta}
 \varepsilon^{\alpha'\beta'\gamma'\delta'}
 D_{k\beta} D_{l\gamma} D_{m\delta}
 D^{(i}_{\beta'} D^j_{\gamma'} A^{klm)}_{\delta'}
 \non\\&&
-\frac1{64512} \frac1{\square^2} \varepsilon^{\alpha\beta\gamma\delta}
\varepsilon^{\alpha'\beta'\gamma'\delta'}
D_{k\alpha} D_{l\beta} D_{m\gamma} D_{n\delta}
 D^{(i}_{\alpha'} D^j_{\beta'} D^k_{\gamma'} A^{lmn)}_{\delta'}
~.
\eea
Note that the terms $\Lambda_2^{++}$, $\Lambda_4^{++}$, $\Lambda_5^{++}$ and $\Lambda_8^{++}$ do not contribute to this expression owing to 
the identity  \cite{GIKOS}
$\int \rd u\, D^{--}F^{++}=0$, for any smooth field $F^{++}(u)$.

%%%%%%%%%%%%%%%%%%%%%%%%%%%%%%%%%%%%%%%%%%%%%%%%%%%%%%

%%%%%%%%%%%%%%%%%%%%%%%%%%%%%%%%%%%%%%%%%%%%%%%%%%%%%%

\begin{footnotesize}

\end{footnotesize}

\end{document}